\documentclass{emulateapj}
\usepackage{graphicx, color}
\usepackage{rotating}

\begin{document}

\shorttitle{$z\simeq6.2$ Quadruply-Lensed Dwarf Galaxy}
\shortauthors{Zitrin et al.}

\slugcomment{Submitted to the Astrophysical Journal Letters}

\title{CLASH: Discovery of a Bright $\lowercase{z}\simeq6.2$ Dwarf Galaxy Quadruply Lensed by MACS J0329.6-0211}

\author{A. Zitrin\altaffilmark{1}}
\author{J. Moustakas\altaffilmark{2}}
\author{L. Bradley\altaffilmark{3}}
\author{D. Coe\altaffilmark{3}}
\author{L.A. Moustakas\altaffilmark{4}}
\author{M. Postman\altaffilmark{3}}
\author{X. Shu\altaffilmark{5}}
\author{W. Zheng\altaffilmark{6}}
\author{N. Ben\'itez\altaffilmark{7}}
\author{R. Bouwens\altaffilmark{8}}
\author{T. Broadhurst\altaffilmark{9,10}}
\author{H. Ford\altaffilmark{6}}
\author{O. Host\altaffilmark{11}}
\author{S. Jouvel\altaffilmark{11}}
\author{A. Koekemoer\altaffilmark{3}}
\author{M. Meneghetti\altaffilmark{12}}
\author{P. Rosati\altaffilmark{13}}
\author{M. Donahue\altaffilmark{14}}
\author{C. Grillo\altaffilmark{15}}
\author{D. Kelson\altaffilmark{16}}
\author{D. Lemze\altaffilmark{6}}
\author{E. Medezinski\altaffilmark{6}}
\author{A. Molino\altaffilmark{7}}
\author{M. Nonino\altaffilmark{17}}
\author{S. Ogaz\altaffilmark{3}}

\altaffiltext{1}{Institut f\"{u}r Theoretische Astrophysik, Universit\"at Heidelberg; adizitrin@gmail.com}
\altaffiltext{2}{Center for Astrophysics and Space Sciences, University of California, San Diego}
\altaffiltext{3}{Space Telescope Science Institute, Baltimore}
\altaffiltext{4}{Jet Propulsion Laboratory, California Institute of Technology}
\altaffiltext{5}{University of Science and Technology of China, Hefei, Anhui}
\altaffiltext{6}{Department of Physics and Astronomy, The Johns Hopkins University}
\altaffiltext{7}{Instituto de Astrof\'isica de Andaluc\'ia (CSIC)}
\altaffiltext{8}{Leiden Observatory, University of Leiden}
\altaffiltext{9}{Department of Theoretical Physics, University of Basque Country}
\altaffiltext{10}{IKERBASQUE, Basque Foundation for Science}
\altaffiltext{11}{Department of Physics \& Astronomy, University College London}
\altaffiltext{12}{INAF, Osservatorio Astronomico di Bologna; INFN, Sezione di Bologna}
\altaffiltext{13}{European Southern Observatory}
\altaffiltext{14}{Physics and Astronomy Department, Michigan State University}
\altaffiltext{15}{Excellence Cluster Universe, Technische Universit\"at M\"unchen}
\altaffiltext{16}{Observatories of the Carnegie Institution of Washington, Pasadena}
\altaffiltext{17}{INAF-Osservatorio Astronomico di Trieste}

\begin{abstract}
We report the discovery of a $z_{phot}=6.18^{+0.05}_{-0.07}$ ($95\%$
confidence level) dwarf galaxy, lensed into four images by the
galaxy cluster MACS J0329.6-0211 ($z_{l}=0.45$). The galaxy is observed as a high-redshift dropout in \emph{HST}/ACS/WFC3
CLASH and \emph{Spitzer}/IRAC imaging. Its redshift is securely determined due to a clear detection of the
Lyman-break in the 18-band photometry, making this galaxy one of the
highest-redshift multiply-lensed objects known to date with an
observed magnitude of F125W$=24.00\pm0.04$ AB $mag$ for its
most magnified image. We also present the first strong-lensing
analysis of this cluster uncovering 15 additional multiply-imaged candidates of five lower-redshift sources spanning the range
$z_{s}\simeq2-4$. The mass model independently supports the high
photometric redshift and reveals magnifications of
$11.6^{+8.9}_{-4.1}$, $17.6^{+6.2}_{-3.9}$, $3.9^{+3.0}_{-1.7}$, and
$3.7^{+1.3}_{-0.2}$, respectively, for the four images of the
high-redshift galaxy. By delensing the most magnified image we construct an image of the source with a
physical resolution of $\sim200$ pc when the universe was $\sim0.9$ Gyr
old, where the $z\simeq6.2$ galaxy occupies a source-plane area of
approximately 2.2 kpc$^{2}$. Modeling the observed spectral energy
distribution (SED) using population synthesis models, we find a demagnified
stellar mass of $\sim10^{9}~\mathcal{M}_{\sun}$, subsolar metallicity
($Z/Z_{\sun}\sim0.5$), low dust content ($A_{V}\sim0.1$~mag), a
demagnified SFR of $\sim3.2~\mathcal{M}_{\sun}$~yr$^{-1}$, and a specific SFR of
$\sim3.4$~Gyr$^{-1}$, all consistent with the properties of local
dwarf galaxies.
\end{abstract}

\keywords{dark matter, galaxies: clusters: individuals: MACS
  J0329.6-0211, galaxies: clusters: general, galaxies: high-redshift,
  gravitational lensing: strong}


\section{Introduction}\label{intro}

By gravitationally lensing distant background sources, galaxy clusters act as natural magnifying lenses
in the sky, providing a unique window into the early Universe. Several high-redshift galaxies have been discovered through the
magnification power of galaxy clusters
\citep[e.g.][]{Franx1997on1358highz, Frye2002z4outflow,
  Bouwens2009behindclusters, Zheng2009, Bradley2011}. The current
highest-redshift lensed galaxy is at $z=7.6$ in the field of
Abell~1689 \citep{Bradley2008}, but among the highest-redshift
\emph{multiply-lensed} galaxies known are a galaxy at $z=6.03$ in the
field of Abell~383 (comprising two images;
\citealt{Richard2011,Zitrin2011c}), a $z\sim6.5$ galaxy in the field
of Abell~2218 (comprising three images;
\citealt{Kneib2004on2218highz,Egami2005on2218highz}), and a $z\sim7$
candidate in the ``bullet cluster'' (comprising two images;
\citealt{Hall2011bullet}). In addition, the lens model can be
used to map the lensed images back into the source plane, while the high magnification
enables their spatially resolved internal structural properties to be
measured. Such measurements are not possible without the aid of the
lensing power of the cluster \citep[e.g.][]{Zitrin2011b}.

\begin{figure*}
\centering
 \includegraphics[width=160mm,trim=0mm 0mm 0mm 0mm,clip]{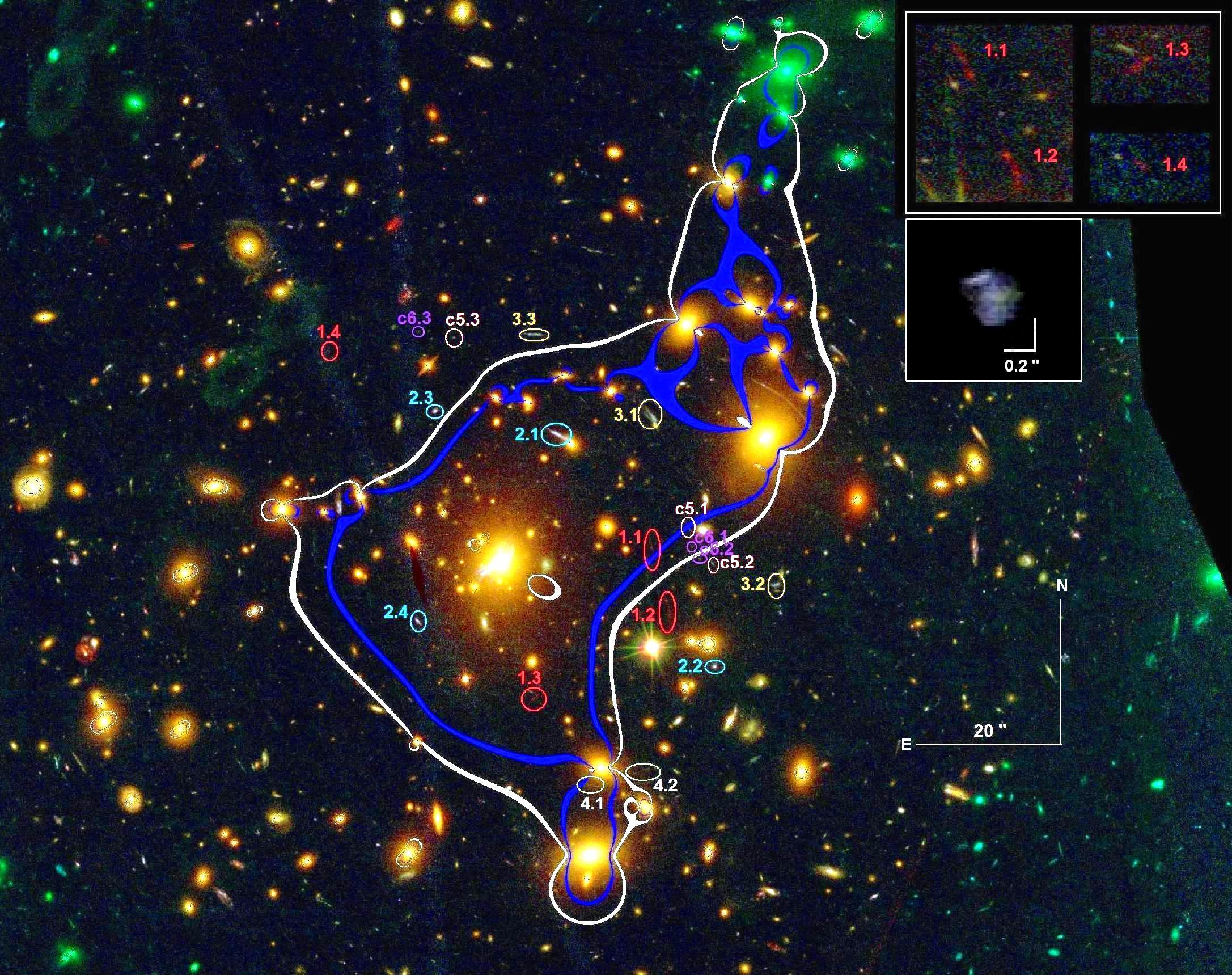}
\caption{Galaxy cluster MACS0329 ($z=0.45$) imaged with
  \emph{HST}/ACS/WFC3. We number the multiple-images uncovered by our model, where the two candidate systems are
  marked in ``c''. All systems are accurately reproduced by
  our model, with an average image-plane reproduction
  uncertainty of $2.07\arcsec$ per image, and image-plane $rms$ of
  $2.32\arcsec$, excluding candidate systems. The white critical curve corresponds to system 1 at $z_{s}=6.18$, and the blue critical curve corresponds to system 2 at $z_{s}=2.17$. The composition of this color image is Red=F105W+F110W+F125W+F140W+F160W, Green=F606W+F625W+F775W+F814W+F850LP, and Blue=F435W+F475W. The
  \emph{upper-right} inset shows a blow-up of the $z=6.18$ system, below which we inset the source-galaxy as reproduced by our model
  by delensing the most magnified image 1.2 ($\mu\sim17.6$) to the
  source plane, yielding a resolution of $\sim200$ pc per pixel. Overplotted therein are bars
  showing the source-plane size of $0.2\arcsec$ ($\simeq1.2$ kpc). Noise cleaning and color manipulation procedures were performed for
  a better view of the internal details of the source.}
\label{curves0329}
\end{figure*}

The Lyman-alpha forest produces a sharp drop in flux below
$1216$~\AA. The expansion of the universe moves this spectral break to
longer wavelengths, allowing high-redshift galaxies to be identified as
``dropouts'' in the observed-frame optical and near-IR \citep[see][]{Madau1995dropout,Franx1997on1358highz}.  As these young,
high-redshift galaxies are usually actively forming stars, their
rest-frame UV spectra should be relatively blue. This combination of properties allows the redshift of the galaxy to be
estimated accurately, even without spectroscopy: as low as $\sim1\%$ uncertainty on the redshift is obtained here, for 95\% confidence levels. The study of high-redshift galaxies enables important constraints to be placed on galaxy evolution and structure formation. Particularly, observing high-redshift galaxies provides direct measurements of the high-$z$ luminosity-function, early SFR, and the epoch of the intergalactic medium reionization \citep[see][and references therein]{Bradley2011}.

We report the discovery of one of the highest-redshift
($z_{phot}=6.18^{+0.05}_{-0.07}$) multiply-lensed galaxies known to date, lensed into four images by the galaxy cluster MACS J0329.6-0211 ($z$=0.45; MACS0329 hereafter). MACS0329 is an X-ray selected system found by the Massive Cluster Survey, MACS
(\citealt{EbelingMacsCat2001,EbelingMacsFull2010}). \citet{Maughan2008xray} classified MACSJ0329 as having evidence for substructure in its X-ray surface brightness, although \citet{SchmidtAllen2007} classified it as relaxed. We found no record
of previous strong lensing (SL) analysis of this cluster, and only 1-band shallow \emph{HST}/WFPC2 previous imaging. The 16
\emph{HST} bands chosen for the Cluster Lensing and Supernova survey with Hubble (CLASH; \citealt{PostmanCLASHoverview}) enable us to conduct the first SL study of
this cluster, and to obtain accurate photometric redshifts for the multiply-lensed
background galaxies. We use these data in conjunction with a parametric SL modeling method \citep[e.g.,][]{Broadhurst2005a, Zitrin2009b, Zitrin2011a, Zitrin2011b, Merten2011}, to find several multiple image families across the central field of MACS0329 (Fig. \ref{curves0329}) so that its mass distribution and inner profile can be
well-constrained, allowing us to deduce the source-plane properties of the high-redshift galaxy presented here.

The paper is organized as follows: In \S 2 we describe the
observations, and in \S 3 we detail the SL model. In \S 4 we report the results of the SL analysis and the physical properties
of the high-redshift galaxy based on detailed SED modeling. The results and conclusions
are summarized in \S 5. We adopt a concordance
$\Lambda$CDM cosmology with $\Omega_{\rm m0}=0.3$, $\Omega_{\Lambda
  0}=0.7$, and $h=0.7$. With these parameters, one arcsecond
corresponds to a physical scale of $5.76$~kpc at the cluster redshift
($z=0.45$; \citealt{Allen2008xrayCLASH}), and $5.62$~kpc at $z=6.18$.

\section{Observations and Redshifts}\label{obs}

\begin{figure*}
\vspace{0.1cm}
 \begin{center}
 \includegraphics[width=160mm,trim=0mm 0mm 0mm 0mm,clip]{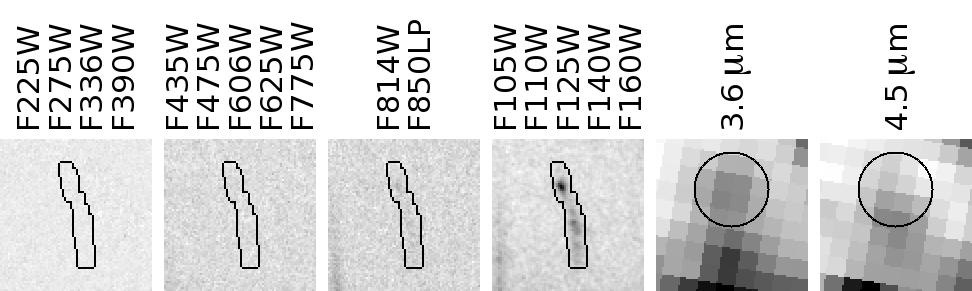}
 \end{center}
\caption{Thumbnails ($5\arcsec \times 5\arcsec$) of the four arcs of the $z=6.18$ source in the different bands, with tailored apertures used for the photometry overlaid. As is clear, these are not seen below the F814W band, in agreement with the photometric redshift estimate. See Fig. \ref{SF} and Table \ref{systems} for the SED and photometry, respectively.}
\label{dropout}
\end{figure*}

MACS0329 was observed as part of CLASH with \emph{HST} from 2011 August to
2011 October.  This is one of 25 clusters to be observed to a total depth of
20 \emph{HST} orbits in 16 filters with the Wide Field Camera 3 (WFC3) UVIS
and IR cameras, and the Advanced Camera for Surveys (ACS) WFC.  The
images are reduced and mosaiced ($0\farcs065$~pixel$^{-1}$) using standard techniques implemented in the {\tt MosaicDrizzle}
pipeline \citep{Koekemoer2002MultiDrizzle,Koekemoer2011Candles}. HST photometry for the arcs is obtained within the apertures shown in Fig. \ref{dropout}, and photometric redshifts are estimated for all the galaxies in the field using the full 16-band UVIS/ACS/WFC3-IR aperture-matched (e.g., Fig \ref{dropout}) photometry via both the BPZ \citep{Benitez2000, Benitez2004, Coe2006} and LePhare \citep{ArnoutsLPZ1999,Ilbert2006BPZ} programs. See \citet{PostmanCLASHoverview} for additional details.

We supplement these \emph{HST} observations with Infrared Array Camera
(IRAC) imaging obtained in 2011 October as part of the \emph{Spitzer} Cycle~8 warm mission (PI: R. Bouwens; see also \citealt{PostmanCLASHoverview}).  The Warm \emph{Spitzer} IRAC observations in Channels 1 and 2 (3.6 and 4.5$\mu$m) have a total integration time of $13,290$s, and clearly detect three of the four images of the background source we report here (1.1 through 1.3), with a formal $\sim2\sigma$ detection for arc 1.4. Because of crowding in the Spitzer images, we adopt a circular $2.4\arcsec$ diameter aperture for the IRAC photometry. Background levels are estimated by sampling non-crowded sections of the IRAC mosaic while avoiding nearby bright sources. The accuracy of this (manual) photometry is verified by subtracting possible contaminating sources with GALFIT\footnote[1]{http://users.obs.carnegiescience.edu/peng/work/galfit/galfit.html}, especially for arc 1.2 which has a bright star nearby. We use a flux aperture correction factor of $1.9\times$, based on a curve of growth analysis of unresolved sources in the Post Basic Calibrated Data (PBCD) mosaic. The relevant multi-wavelength photometry is given in Table \ref{systems}.

\section{Strong Lensing Modeling and Analysis}\label{model}

The lens modeling method used here \citep[e.g.,][]{Broadhurst2005a, Zitrin2009b} begins with the assumption that mass approximately traces light. We model the distribution of cluster mass by assigning a power-law mass profile to each red-sequence cluster galaxy, scaled by its (relative) brightness. The sum of all galaxy contributions represents the lumpy galaxy component, which is then smoothed using 2D spline interpolation to obtain a smooth-component representing the dark matter (DM) distribution. The polynomial degree of smoothing and the index of the power-law are the most
important free parameters determining the mass profile.

A worthwhile improvement in fitting the location of the lensed images is generally found by
introducing an external shear describing the overall
matter ellipticity. The direction of the shear and its amplitude are
free parameters, allowing for some flexibility in the relation between
the distribution of DM and the distribution of galaxies. The weight of the lumpy component relative to the DM, and the overall normalization, bring the total number of free parameters in our modeling to six \citep[see][]{Zitrin2009b}.

\begin{figure}
\vspace{0.5cm}
 \begin{center}
  \includegraphics[width=80mm]{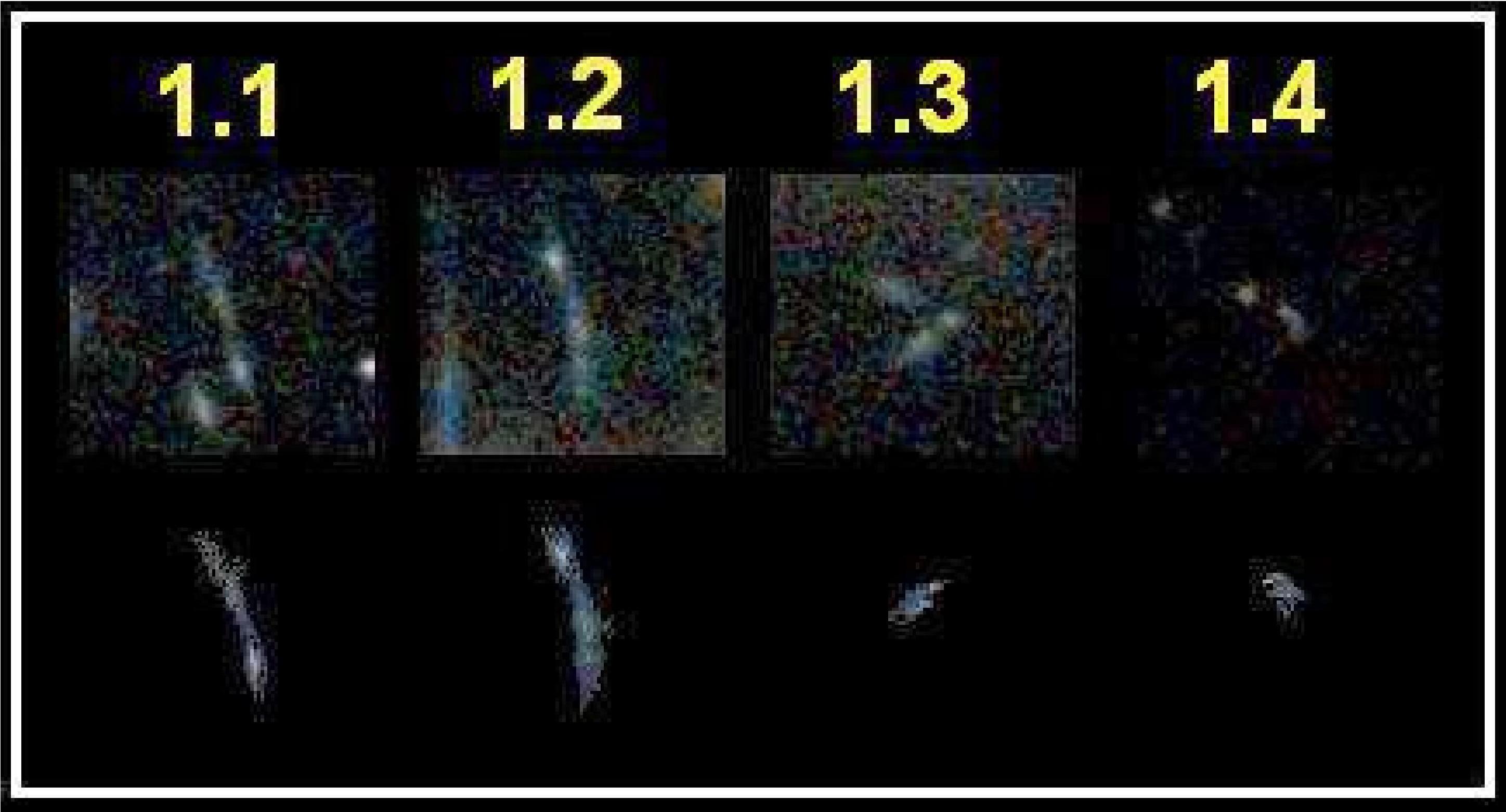}
 \end{center}
\caption{Reproduction of systems 1 by our model (\emph{lower row}), compared with the real images (\emph{upper row}). We delens image 1.1 to the source plane and relens it back to the image-plane to reproduce the other images of this system. As can be seen, our model reproduces well the images of this system. Note also, our model requires larger deflection angles for the $z=6.18$ source, relative to lower-$z$ systems such as systems 2 and 3, independently strengthening its detection as a high-redshift source.}
\label{Rep1to4}
\end{figure}

The best fit is assessed by a $\chi^{2}$ minimization in the image plane, which is preferred because it generally does not produce solutions that are biased towards shallow mass profiles (and high magnifications), as is the case with source-plane minimization. The positions and morphologies of the multiply-imaged arcs are accurately reproduced by the best-fit mass model in their measured photometric redshifts (for a morphological comparison example see Fig. \ref{Rep1to4}), and the model is successively refined as additional sets of multiple images are incorporated.

\begin{table*}
\caption{Multiple-images data and multiband photometry for the four images of the $z=6.18$ galaxy}
\label{systems}
\begin{center}
\scriptsize\begin{tabular}{|c|c|c|c|c|c|c|c|c|c|c|}
\hline\hline
ARC & RA & DEC & \multicolumn{3}{c|}{BPZ $z_{phot}$}& \multicolumn{3}{c|}{LPZ $z_{phot}$}& \multicolumn{2}{c|}{Magnification, $\mu$} \\
ID& (J2000.0)&(J2000.0)& \multicolumn{3}{c|}{(best) [95\% C.L.]}& \multicolumn{3}{c|}{(best) [95\% C.L.]} &\multicolumn{2}{c|}{} \\
\hline
1.1 & 03:29:40.18 & -02:11:45.60 & \multicolumn{3}{c|}{6.185 [6.079--6.282]}& \multicolumn{3}{c|}{6.155 [5.885--6.374]} &\multicolumn{2}{c|}{$11.6^{+8.9}_{-4.1}$}\\
1.2 & 03:29:40.06 & -02:11:51.72 & \multicolumn{3}{c|}{6.137 [6.024--6.204]}& \multicolumn{3}{c|}{6.194 [5.982--6.347]} & \multicolumn{2}{c|}{$17.6^{+6.2}_{-3.9}$} \\
1.3 & 03:29:41.28 & -02:12:05.04 & \multicolumn{3}{c|}{6.152 [6.055--6.257]}& \multicolumn{3}{c|}{6.253 [5.828--6.573]} & \multicolumn{2}{c|}{$3.9^{+3.0}_{-1.7}$} \\
1.4 & 03:29:43.16 & -02:11:17.30 & \multicolumn{3}{c|}{6.246 [6.073--6.395]}& \multicolumn{3}{c|}{5.965 [5.482--6.327]} & \multicolumn{2}{c|}{$3.7^{+1.3}_{-0.2}$}\\
\hline
\multicolumn{11}{|c|}{Multi-band photometry}\\
\hline
ARC & F775W & F814W & F850LP & F105W & F110W & F125W & F140W & F160W & $3.6~\mu m$ & $4.5~\mu m$\\
\hline
1.1 &  $26.24\pm0.32$ & $25.69\pm0.12$ & $24.73\pm0.11$ & $24.41\pm0.05$ & $24.39\pm0.04$ & $24.46\pm0.05$  & $24.52\pm0.04$  & $24.62\pm0.05$  & $24.4\pm0.2$ &   $24.7\pm 0.3$\\
\hline
1.2 & $>26.52$~(2$\sigma$) & $25.32\pm0.10$ & $24.29\pm0.08$ & $23.94\pm0.04$ & $23.90\pm0.03$  & $24.00\pm0.04$ & $24.04\pm0.03$ & $24.11\pm0.04$  & $23.6\pm0.2$ & $23.5\pm0.1$\\
\hline
1.3 & $26.52\pm0.33$ & $26.30\pm 0.17$ & $25.52\pm0.17$ & $24.83\pm0.06$ & $24.93\pm0.05$ & $24.87\pm0.06$ & $24.91\pm0.05$ & $25.01\pm0.06$ & $24.9\pm$0.6& $24.5\pm$0.3\\
\hline
1.4 & $>29.94$~(2$\sigma$) & $>27.76$~(2$\sigma$) & $25.28\pm0.14$ & $26.17\pm0.13$ & $26.25\pm0.10$ & $26.54\pm0.17$ & $26.42\pm0.13$ & $26.28\pm0.13$ & $24.8\pm0.2$ & $>24.9$~(2$\sigma$)\\
\hline
\end{tabular}
\tablecomments{Data for the multiple images of the $z=6.18$ galaxy. \emph{Top:} Columns are: arc ID; RA and DEC; photo-$z$ using BPZ and LePhare (hereafter LPZ), respectively; $\mu$, magnification predicted by the mass model. \emph{Bottom:} Multiband photometry for the dropout band (F775W) and all redder bandpasses, for the four images of the $z=6.18$ galaxy, in AB mags. Lower ($2\sigma$) limits are given for non-detections; we find no $2\sigma$ detections in any of the 8 bluer bands. Note also that the (relative) magnitudes between the different images of this system agree overall with the (relative) magnifications derived by our mass model (see \S \ref{results}), with some deviation for the less magnified images, 1.3(1.4), which seem slightly less(more) magnified than our model implies.}
\end{center}
\end{table*}

\section{Results}\label{results}

We uncovered in \emph{HST}/CLASH and \emph{Spitzer}/IRAC imaging
of MACS0329 one of the highest-redshift multiply-lensed galaxies known
to date. We use the independent photometric-redshift distributions of
its four lensed images to obtain a source redshift (and $2\sigma$
limits, see Table 1) of $z_{phot}=6.18^{+0.05}_{-0.07}$. Using our mass model and
the extensive imaging and resulting photometric redshifts, we physically matched 15 additional new multiple images and candidates of
five background, lower-$z$ sources (see Figure \ref{curves0329}), which are used in turn to refine the mass model. Explicitly, in addition to the four images of the high-$z$ galaxy, we use as additional constraints the seven multiple-images of systems 2 and 3, at photometric redshifts (and $2\sigma$ limits) of $z_{s}=2.17^{+0.12}_{-0.03}$ and $z_{s}=2.89^{+0.05}_{-0.05}$, respectively, and verify that all other systems and candidates are plausible in the context of the resulting mass model.

The magnification by the cluster is coupled to the surface-density
gradient in each point (i.e., the mass profile), so
multiple-systems with redshift information are important to assess the
true magnification across the cluster field. For the four images of
the $z_{phot}=6.18$ galaxy, we obtain magnifications (and $1\sigma$
uncertainties) of $11.6^{+8.9}_{-4.1}$, $17.6^{+6.2}_{-3.9}$,
$3.9^{+3.0}_{-1.7}$, and $3.7^{+1.3}_{-0.2}$, respectively, so that in
total the galaxy is magnified by a factor of $\simeq37$.

\begin{figure}
\vspace{0.1cm}
 \begin{center}
 \includegraphics[width=82mm,trim=0mm 0mm 0mm 0mm,clip]{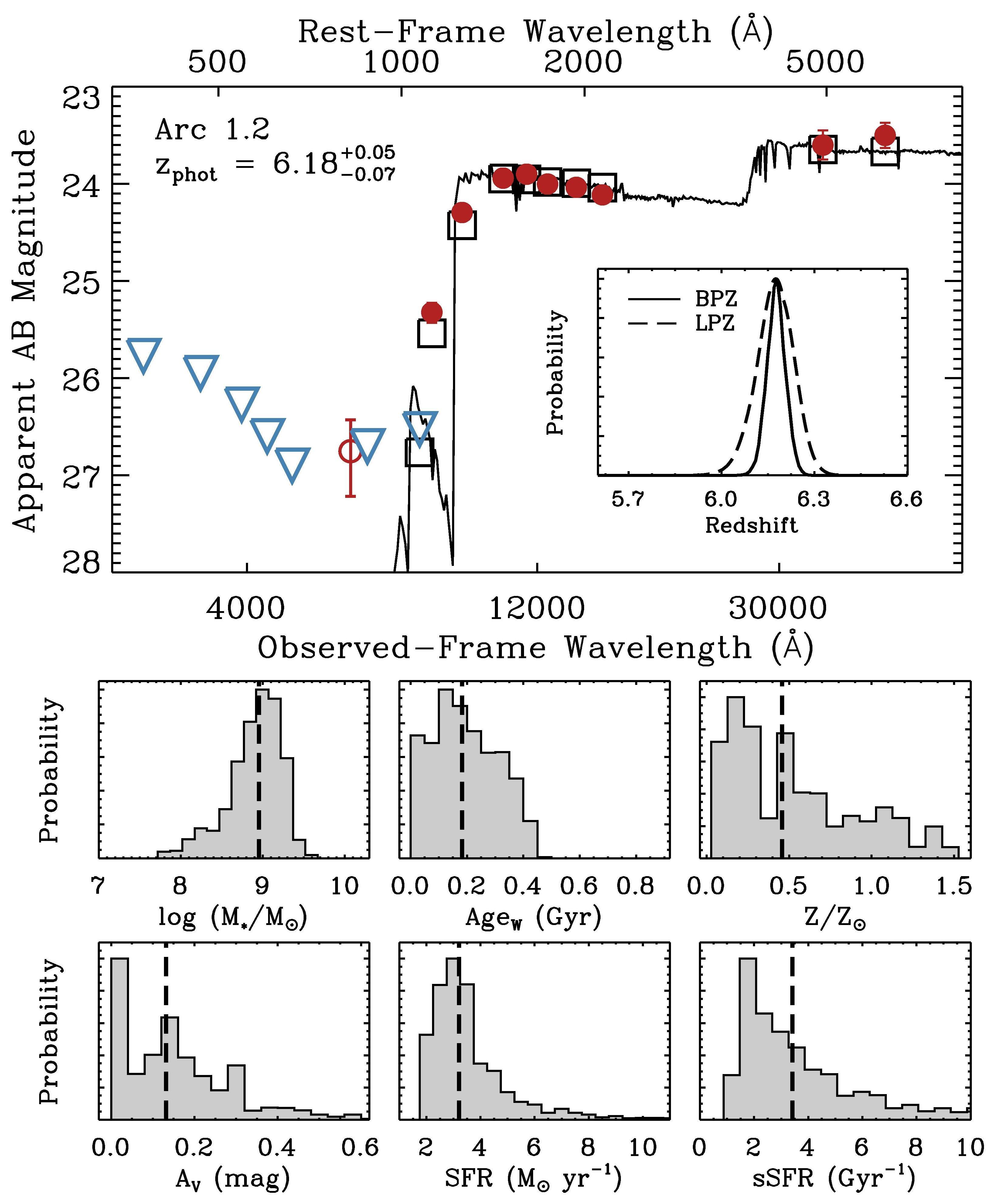}
 \end{center}
\caption{SED modeling results of arc~1.2, the most magnified
  image of the $z_{phot}\simeq6.18$ galaxy.  The upper panel shows the
  observed and rest-frame SED and the maximum likelihood model
  fit. The filled red points indicate bands in which the object is
  well detected, while the open blue triangles indicate $2\sigma$
  upper limits; the open squares show the photometry of the
  best-fitting model convolved with the WFC3, ACS, and IRAC filter
  response functions.  The open red circle indicates a spurious,
  marginally significant detection of arc~1.2 in the F606W band;
  however, we emphasize that none of the other images are detected in
  this band.  We also show an inset with the photometric redshift
  distribution, $P(z)$, obtained by combining the independent
  photometric redshift measurements of the four arcs, yielding a
  source redshift of $z=6.18^{+0.05}_{-0.07}$ with BPZ, and
  $6.18^{+0.11}_{-0.13}$ with LPZ ($95\%$ confidence levels). The lower panels show the posterior
  distributions on the demagnified stellar mass, SFR weighted age, stellar metallicity (relative to
  $Z_{\sun}=0.019$), rest $V$-band dust attenuation, demagnified SFR,
  and specific SFR, sSFR$\equiv$SFR/$\mathcal{M}$.  The vertical
  dashed lines show the median of each posterior distribution.  We
  summarize the results in \S \ref{results}.}
\label{SF}
\end{figure}

We leverage the magnification boost of the most magnified image, arc
1.2, to construct a high-resolution image of the background galaxy, which occupies a source-plane area of $\sim2.2$ kpc$^{2}$ (see Figure
\ref{curves0329}). To further constrain the physical properties of this galaxy, we model
the observed SED of arc 1.2 using the Bayesian SED-fitting code {\tt
  iSEDfit} (see \citealt{Moustakas2011inprep}) coupled to the flexible stellar population synthesis models of
\citet{Conroy2009SFS}. We adopt the \citet{Chabrier2003IMF} initial
mass function from $0.1-100~\mathcal{M}_{\sun}$, assume the
\citet{Calzetti2000dust} dust attenuation law, and adopt uniform
priors (based on Monte Carlo draws) on the stellar metallicity
($0.0002<Z<0.03$), $V$-band attenuation ($0<A_{V}<2$~mag), and galaxy
age ($0.005-1$~Gyr). For reference, the age of the universe at
$z=6.18$ is $0.9$~Gyr. We parameterize the star formation history
$\psi(t)$ as an exponentially declining function of time, $t$, given
by $\psi(t)\propto\exp(-t/\tau)$, where $\tau$ is the characteristic
time for star formation.  We draw $\tau$ from a uniform distribution
between $0.01-5$~Gyr, which spans the range from passively evolving to
continuous star formation.

In Figure~\ref{SF} we show the observed and rest-frame SED of arc~1.2,
the maximum likelihood model fit, and the posterior distributions on
all the model parameters. Adopting the median of each posterior
distribution as our best estimate of the properties of the galaxy, we
find a demagnified stellar mass of $\sim10^{9}~\mathcal{M}_{\sun}$,
low dust content ($A_{V}\sim0.1$~mag), a demagnified SFR of $\sim3.2~\mathcal{M}_{\sun}$~yr$^{-1}$, and a
SFR-weighted age of $\sim180$~Myr. These results imply a specific
SFR, sSFR$\equiv$SFR/$\mathcal{M}$, of $\sim3.4$~Gyr$^{-1}$,
corresponding to a mass-doubling time of just $600$~Myr (assuming a 50\% return fraction). The stellar
metallicity and $\tau$ parameter (not shown) are not particularly well
constrained, although solutions with subsolar metallicity and
$\tau\gg0$~Gyr are generally favored; the median of the posterior
distributions imply subsolar metallicity, $Z/Z_{\sun}\sim0.5$, and
$\tau\sim2.4$~Gyr.

We verified that performing our SED modeling on the other (less
magnified) images yields overall similar results. The demagnified SED of arc 1.1
is nearly identical to that of arc 1.2, and therefore the posterior
distributions of the physical quantities we derive are very similar to
those shown in Figure \ref{SF}; the median quantities all agree to well
within ~0.1 dex ($\pm50\%$). The demagnified F125W magnitude of arc
1.3(1.4), on the other hand, is ~0.8 mag brighter(fainter) than arcs
1.1 and 1.2, although this discrepancy is well within the statistical
uncertainties on the magnifications. Nevertheless, at face
value the fainter two arcs imply an 0.3-0.4 dex larger stellar mass
and SFR, 0.2-0.6 mag more dust attenuation, and a very similar age,
~200 Myr. However, we emphasize that the posterior distributions on
these quantities overlap significantly with those of arcs 1.1 and 1.2,
and therefore these differences do not affect our conclusions about
the nature of this object.

Note also that the center of emission in the 3.6$\mu$m image is slightly different than that of the 4.5$\mu$m image (Figure \ref{dropout}). Repeating the photometry on the two different centers, we find this offset may introduce uncertainty of $\sim$0.3 mag in the IRAC photometry. The effect of possible contamination from bright neighbors was re-examined for both emission-centers, and was found to be typically 0.4 mag. However, we importantly verified that these higher uncertainties have only a negligible ($<0.1$ dex) effect on the resulting physical properties, as the fit is governed by the HST photometry.

Finally, we use our lens model to constrain the physical
characteristics of MACS0329.  For the $z_{s}=6.18$ source,
the critical curves enclose a relatively large area, with an effective
Einstein radius of $r_{E}=33\farcs9\pm3\arcsec$ ($\simeq$195 kpc at
$z_{l}=0.45$), and a projected mass of $1.89^{+0.10}_{-0.06}
\times 10^{14}~\mathcal{M}_{\sun}$ (Figure \ref{curves0329}). For the lower redshift of system 2, $z_s=2.17$, the Einstein radius is
$\simeq27\farcs7$, and the critical curve encloses a projected mass of $1.40^{+0.09}_{-0.05}
\times10^{14}~\mathcal{M}_{\sun}$. For completeness, we measure the (total) mass profile slope, $d\log
\Sigma/d\log r\simeq -0.61^{+0.05}_{-0.1}$ (in the range
$1\arcsec-52\arcsec$, or $6\la r \la300$ kpc; about twice the Einstein
radius for $z_{s}\sim2$), typical of \emph{relaxed} and well-concentrated lensing
clusters \citep[e.g.][]{Broadhurst2005a,Zitrin2009b}, and in agreement
with the fairly circularly-symmetric X-ray contours centered on the
BCG \citep[see][]{MannEbeling2011}.

\section{Discussion and Conclusions}\label{conclusions}

The discovery of a high-redshift galaxy in the field of MACS0329 adds
to several known high-reshift galaxies lensed by galaxy clusters
\citep[e.g.,][]{Egami2005on2218highz,Bradley2008,Bradley2011,Zheng2009,Richard2011}. Here we summarize
the properties of this unique source.

(1) It is one of the highest-redshift multiply lensed objects known to
date, and lensed into four separate images. The angular separation
between arcs 1.2 and 1.4 is $\sim1\arcmin$, considerably larger than
previously reported cases \citep{Egami2005on2218highz, Richard2011}.

(2) The source is one of the brightest at $z>6$:  its $J_{125}$
magnitude is 24.0 AB, making it a viable candidate for follow-up
spectroscopy.

(3) The galaxy is consistent with being a dwarf galaxy. Its intrinsic
(delensed) magnitude of $J_{125} = 27.1$ AB makes it a sub-$L_{*}$
galaxy at this redshift.  It occupies a source-plane area of
$\sim$2.2 kpc$^{2}$, similar to previously deduced sizes of
high-$z$ lensed galaxies \citep[e.g.][]{Zitrin2011b}. Due to the
hierarchical growth of structure, galaxies are expected to be small at
high redshifts, with dwarf galaxies constituting the building material
of larger structures. Our source-plane reconstruction shows at least three
(possibly star-forming) knots, consistent with several other reports of
high-redshift galaxies with multiple components
\citep{Franx1997on1358highz,Bradley2008,Bradley2011, Zheng2009,Oesch2010highz,Zitrin2011b}, possibly as
the result of merging. In addition, we measure an overall
half-light radius of $\sim0.12\arcsec$, consistent with that found in
\citet{Bouwens2004size,Bouwens2006size} and \citep{Oesch2010highz}.

(4) The SED fits to the multiband photometry of the source suggest a
demagnified stellar mass of $\sim10^{9}~\mathcal{M}_{\sun}$, a
SFR-weighted age of $\sim180$~Myr, subsolar metallicity
($Z/Z_{\sun}\sim0.5$), low dust content ($A_{V}\sim0.1$~mag), and a
demagnified SFR of $\sim3.2~\mathcal{M}_{\sun}$~yr$^{-1}$.  The specific
SFR of $\sim3.4$ Gyr$^{-1}$, which is slightly higher than that found
by other recent studies \citep{Gonzalez2011highz,Stark2009highz,Labbe2010highz,McLure2011MNRAShighz}, implies a
mass-doubling time of just $600$~Myr and therefore vigorous ongoing
star formation considering its low mass.

(5) The UV continuum is blue, with a UV-slope $\beta = -2.5 \pm 0.06$, consistent with measurements of other faint $z\sim6$ galaxies
and suggests that these sources are largely dust free \citep{Bouwens2009uvslopes,Bouwens2011uvslope,Finkelstein2011highzUV,Vanzella2011uvslope}.

The discovery of the galaxy presented here shows once more the novelty
and tremendous potential of galaxy clusters for observationally
accessing the faint early universe.

\section*{acknowledgments}

The authors thank Saurabh Jha for useful discussions, and the anonymous referee for valuable comments. The CLASH Multi-Cycle Treasury Program (GO-12065) is based on observations made with the NASA/ESA Hubble Space Telescope. The Space Telescope Science Institute is operated by the Association of Universities for Research in Astronomy, Inc. under NASA contract NAS 5-26555. This work is based in part on observations made with the \emph{Spitzer} Space Telescope, which is operated by the Jet Propulsion Laboratory, California institute of Technology under a contract with NASA. Support for this work was provided by NASA through an award issued by JPL/Caltech. We thank the hospitality of Institut f\"{u}r Theoretische Astrophysik, Universit\"at Heidelberg, where part of this work took place, and the Baden-W\"uerttemberg Stiftung.


\end{document}